# 1.5μm Polarization-Entangled Bell States Generation Based on Birefringence in High Nonlinear Microstructure Fiber

**Qiang Zhou[1,2], Wei Zhang[1,3], Jierong Cheng[1], Yidong Huang[1], Jiangde Peng[1]**

[1] Department of Electronic Engineering, Tsinghua University Beijing, 100084, P. R. China

[2] Email: betterchou@gmail.com [3] Corresponding author: zwei@mail.tsinghua.edu.cn

Abstract: Polarization-entangled photon pair generation based on two scalar scattering processes of the vector four photon scattering has been demonstrated experimentally in high nonlinear microstructure fiber with birefringence. By controlling the pump polarization state, polarization-entangled Bell states can be realized. It is provides a simple way to realize efficient and compact fiber based polarization-entangled photon pair sources.

1.5μm polarization-entangled photon pair sources are important devices for quantum communication [1] and quantum information processing [2]. In recent years, spontaneous four-photon scattering (FPS) in optical fibers focuses much attention as a way to generate correlated/entangled photon pairs for its potential in realizing efficient and compact all fiber 1.5μm sources. To realize polarization entanglement in fibers, two schemes have been proposed, based on the time-multiplexing [3] and the polarization diversity loop [4], respectively.

Recently, high nonlinear microstructure fibers (HN-MSFs) have been looked as important candidates for correlated/entangled photon pair generation. Comparing with dispersion shifted fibers used in previous works, the fiber length can be shortened to several meters [5, 6] thanks to their high nonlinearity. Utilizing their unique dispersion characteristic, the generation of correlated/entangled photon pairs can be extended to 800nm band [6] and 800nm/1.5μm band [7, 8].

On the other hand, non-ignorable birefringence is an intrinsic characteristic of HN-MSFs [9]. Recently, the vector FPS theory in fibers with birefringence [10] shows that two kinds of FPS processes take place simultaneously when pump light passes through birefringent fiber. One is scalar scattering process, in which two annihilated pump photons and the generated signal and idler photons are polarized along the same fiber polarization axis. The other is vector scattering process, in which the pump photons, signal and idler photons are polarized along two different fiber polarization axes. Frequencies of the generated signal and idler photons through scalar and vector scattering processes are different, which provide the feasibility to realize polarization-entangled photon pair generation directly in fibers with birefringence.

In this paper, we experimentally demonstrate that polarization-entangled photon pairs can be generated through vector FPS in HN-MSFs with birefringence and polarization-entangled Bell states can be realized by controlling the pump polarization state.

The experiment setup is shown in Fig.1. The HN-MSF is 25m long, fabricated by Crystal fiber A/S Inc. (its SEM image is shown in Fig.1). The diameters of the silica core and air holes are 1.8μm and 0.89μm, respectively. One of its zero dispersion wavelengths is at 1564.8nm. It has an ultra-high nonlinear coefficient of 66.7/W/km, a phase birefringence of $\Delta n=3.5\times10^{-5}$ and a group birefringence of $\Delta\beta_1= 0.158$ps/m at 1.5μm [11, 12].

The pulsed pump light is generated by a passive mode locked fiber laser, whose center wavelength, spectral width and repeating frequency are 1552.75nm, 0.2nm and 1MHz, respectively. A side-band suppression of 115dB is realized by a filter system based on fiber Bragg gratings (FBGs), circulators (Cs) and tunable optical band-pass filters (TOBFs). The power and polarization state of pump light are controlled by a variable optical attenuator (VOA1), a polarizer (P) followed by a rotatable half wavelength plate (HWP1) and a polarization controller (PC1), respectively. The generated correlated photon pairs are directed into two single photon detectors (SPDs) through a filtering and splitting system based on FBGs, circulators and TOBFs. The center wavelength and spectral width of the selected signal photons are 1555.6nm and 0.2nm, while, 1549.9nm and 0.2nm for the idler ones. The total pump suppression is >105dB for either signal or idler wavelengths. The two SPDs (Id201, Id Quantique) are operated in gated Geiger mode with a 2.5ns detection window, triggered with residual pump light detected by a photon detector (PD).

The pump polarization dependence of the correlated photon pair generation in the HN-MSF under linearly polarized pump has been demonstrated in our previous work [13]. It shows that the measured correlated photon pairs are generated by two independent scalar scattering processes in different fiber polarization axes shown in Fig.1, and the impact of polarization dependent dispersion of the HN-MSF can be neglected in the experiment. Hence, polarization entanglement could be expected if the pump polarization direction is not along the fiber axis, which can be realized by rotating the half-wave plate (HWP1) before the HN-MSF. When pump polarization direction is at 45 degree to the fiber polarization axes, the generated polarization-entangled state has the form of $\frac{1}{\sqrt{2}}\left(|H_s\rangle|H_i\rangle + e^{i\phi}|V_s\rangle|V_i\rangle\right)$, where $H$ and $V$ denote the two fiber polarization axes. $s$ and $i$ denote the signal and idler. $\phi$ is the relative phase difference between $|H_s\rangle|H_i\rangle$ and $|V_s\rangle|V_i\rangle$.

In this experiment, in order to demonstrate the polarization entanglement of the generated photon pairs experimentally, two polarization analyzers are inserted before the two SPDs, shown in the dashed square in Fig.1. Each of them consists of a PC, a rotatable HWP and a polarization beam splitter (PBS). Firstly, the polarization direction of linearly polarized pump is adjusted to $H$ axis of the HN-MSF. Hence, the generated correlated photon pairs and spontaneous Raman scattering (SpRS) photons are all polarized along $H$ axis. By adjusting PC2 to achieve maximum signal side count, the signal side polarization analyzer can be collimated to $H$ axis. Rotating HWP2, the signal side detecting polarization direction can be rotated to any angle, denoted by $\theta_s$. Similarly, the idler side detecting direction is denoted by $\theta_i$, which is collimated through PC3 and rotated by HWP3.

Then, polarization direction of the linearly polarized pump is adjusted to 45 degree to the $H$ axis and the polarization-entangled characteristic of generated photon pairs is demonstrated and shown in Fig.2. Fig.2 (a) is the coincidence count per 30s (accidental coincidence count has been subtracted) under different $\theta_i$. The squares and circles are the experiment data when $\theta_s$ is set to 135 and 0 degree, respectively. The solid and dashed lines are fitting curves, showing that the fringe visibilities of two photon interference are 96±3% and 87±4% for $\theta_s$=0 and 135 degree, respectively.

Fig.2 (b) is the idler side photon count under different $\theta_i$. The squares in the inset are the measured results, which vary with $\theta_i$ due to the generated SpRS photons co-polarized with pump light [14]. In our experiment, the pump level is low enough to ensure that the average number of generated photon per pulse is <<1 (typically 0.01 per pump pulse). Hence, the single side counting rates of signal and idler photons (denoted by $N_s$ and $N_i$ respectively) and the coincidence and accidental coincidence counting rates (denoted by $N_{co}$ and $N_{ac}$ respectively) can be expressed as,

$$
\begin{aligned}
N_s &= \eta_s(R + R_s) + d_s \\
N_i &= \eta_i(R + R_i) + d_i \\
N_{co} &= \eta_s\eta_i(R + RR_s + RR_i + R_sR_i) + \eta_s(R + R_s)d_i + \eta_i(R + R_i)d_s \\
N_{ac} &= \eta_s\eta_i(R^2 + RR_s + RR_i + R_sR_i) + \eta_s(R + R_s)d_i + \eta_i(R + R_i)d_s
\end{aligned}
\tag{1}
$$

Where, $\eta_s$ and $\eta_i$ are the collection efficiencies of signal and idler photons. $d_s$ and $d_i$ are the dark count rates of SPDs at signal and idler sides. These parameters have been measured during the experiment setup preparation ($\eta_s$=3.36%, $\eta_i$=2.38%, $d_s=5.98\times10^{-5}\pm1.001\times10^{-6}$ and $d_i=4.67\times10^{-5}\pm0.999\times10^{-6}$). $R$, $R_s$ and $R_i$ are the generation rates of the photon pairs and the SpRS photons at signal and idler wavelengths, respectively. The difference between $N_{co}$ and $N_{ac}$ in the equation (1) results from that they are coincidence count of signal and idler photons generated from the same pump pulse and different pump pulses, respectively. In this experiment, we measure all the $N_s$, $N_i$, $N_{co}$ and $N_{ac}$ under each $\theta_i$ ($\theta_s$=0 during the measurement) and calculate the SpRS contribution to the idler side photon count by solving the equation (1). The dots in the inset of Fig.2 (b) are the calculated SpRS contribution, shown that it is the main contribution of the idler side photon count variation. The main figure of Fig.2 (b) shows the contribution of generated photon pairs to the idler side photon count. It does not vary with $\theta_i$, demonstrating the polarization indistinguishable property at single side of the generated photon pair.

Fig.2 demonstrates that the generated photon pairs are in the polarization-entangled state of $\frac{1}{\sqrt{2}}\left(|H_s\rangle|H_i\rangle + e^{i\phi}|V_s\rangle|V_i\rangle\right)$ under a linearly polarized pump whose polarization direction

is 45 degree to the $H$ axis. However, $\phi$ is unknown in the experiment, which limits its application as a polarization-entangled photon pair source. $\phi$ can be expressed as $\phi=2\phi_p+\phi_b$ in our experiment. $\phi_b$ is fixed phase difference induced by fiber birefringence and $\phi_p$ is phase difference determined by the initial pump polarization state. Hence, $\phi$ can be controlled by adjusting the pump polarization state. Especially, polarization-entangled Bell states $\Psi^{\pm}=\frac{1}{\sqrt{2}}\left(|H_s\rangle|H_i\rangle\pm|V_s\rangle|V_i\rangle\right)$ can be realized while $\phi$ is set to 2nπ and 2nπ+π (n is an integer).

To control $\phi_p$ in our experiment, the pump light is adjusted to elliptically polarized by controlling PC1 and monitored by a polarization analyzer (Santec PEM-320) before it is injected into the HN-MSF. The long elliptical axis is set to 45 degree to the fiber polarization axes, while, the power of the pump light is unchanged. $\phi_p$ can be calculated by $\phi_p=\pm tan^{-1}(10^{-PER/10})$, $PER$ is the measured pump polarization extinction ratio, + and – correspond to the right and left handed elliptically polarized pump light, respectively.

In the experiment, we measure the coincidence count (accidental coincidence count has been subtracted) under different pump $PER$ when $\theta_s$=135 and $\theta_i$=45 degree. The normalized coincidence count rate of the photon pairs, which is denoted by $R_c$, can be calculated by [3]

$$R_c=sin^2\theta_s cos^2\theta_i+cos^2\theta_s sin^2\theta_i+2cos\phi sin\theta_s cos\theta_s sin\theta_i cos\theta_i=0.5[1-cos(2\phi_p+\phi_b)] \qquad (2)$$

By fitting the experiment data according to (2) and the relation between $\phi_p$ and $PER$, $\phi_b$ can be determined.

Fig.3 shows the experiment results of polarization-entangled Bell states generation. The inset is the coincidence count under different pump $PER$s, while the solid curve is the fitting curve according to (2). It can be calculated that $\phi_b$=0.23π in our experiment, hence $\phi$ under each *PER* is determined. The main figure is the coincidence count plotted under different $\phi$, which agrees well with (2). It can be seen that the maximum of the coincidence count indicates $\phi=\pi$, leading to $\Psi^-$ state, while the minimum of the coincidence count indicates $\phi=0$, leading to $\Psi^+$ state. A theoretical two photon interference visibility of 87.5% can be estimated under linearly polarized pump light, which agrees well with experiment results in Fig.2 (a).

In summary, we demonstrate that based on two scalar scattering processes of the vector FPS in HN-MSFs with birefringence, polarization-entangled photon pairs can be generated directly. The polarization-entangled Bell states can be realized by controlling the pump polarization state. It provides a simple way to realize efficient and compact fiber based polarization-entangled photon pair sources.

This work was supported by National Natural Science Foundation of China under Grant No. 60777032, the National Basic Research Program of China (973) under the Grant No. 2010CB327600.

Reference

1. N. Gisin, G. Ribordy., W. Tittel, and H. Zbinden, Rev. Mod. Phys. 74, 145(2002).

2. C. H. Bennett, and S. J. Wiesner, Phys. Rev. Lett. 69, 2881(1992).

3. X. Li, P. Voss, J. E. Sharping, J. Chen, and P. Kumar, Phys. Rev. Lett. 94, 053601(2005).

4. H. Takesue, and K. Inoue, Phys. Rev. A 70, 031802(2004).

5. J. Fan, A. Migdall, and L. J. Wang, Opt. Lett. 30, 3368(2005).

6. J. Fulconis, O. Alibart, W. Wadsworth, P. Russell, and J. Rarity, Opt. Express 13, 7572(2005).

7. E. A. Goldschmidt, M. D. Eisaman, J. Fan, S. V. Polyakov, and A. Migdall, Phys. Rev. A 78, 013844(2008).

8. A. R. McMillan, J. Fulconis, M. Halder, C. Xiong, J. G. Rarity, and W. J. Wadsworth, Opt. Express 17, 6156-6165 (2009).

9. P. Russell, J. Lightw. Technol. 24, 4729(2006).

10. E. Brainis, Phys. Rev. A 79, 023840( 2009).

11.Zhang Wei, Xiao Li, Zhang Lei, Huang Yi-Dong and Peng Jiang-De, Chin. Phys. Lett. 23, 1201(2006).

12. Xiao Li, Zhang Wei, Huang Yi-Dong, and Peng Jiang-De, Chin. Phys. B .17, 995(2008).

13.Q. Zhou, W. Zhang, S. Zhang, J. Cheng, Y. Huang, J. Peng, OFC'09 Conference, OWD6(2009).

14. Q. Lin, F.Y., and G. P. Agrawal, Phys. Rev. A 75, 023803(2007).

Figure Captions

Fig. 1 (Color online) Experiment setup

VOA: Variable Optical Attenuator, P: Polarizer, HWP: Half Wavelength Plate; PC: Polarization Controller, C: Circulator; FBG: Fiber Bragg Grating; TOBF: Tunable Optical Band-pass Filter; PBS: Polarization Beam Splitter; PD: Photon Detector; SPD: Single Photon Detector

Fig. 2 (Color online) Experiment results of polarization entanglement demonstration (a) coincidence count under different $\theta_i$, (b) idler side photon count under different $\theta_i$.

Fig. 3 (Color online) Polarization-entangled Bell states generation by controlling the pump polarization state

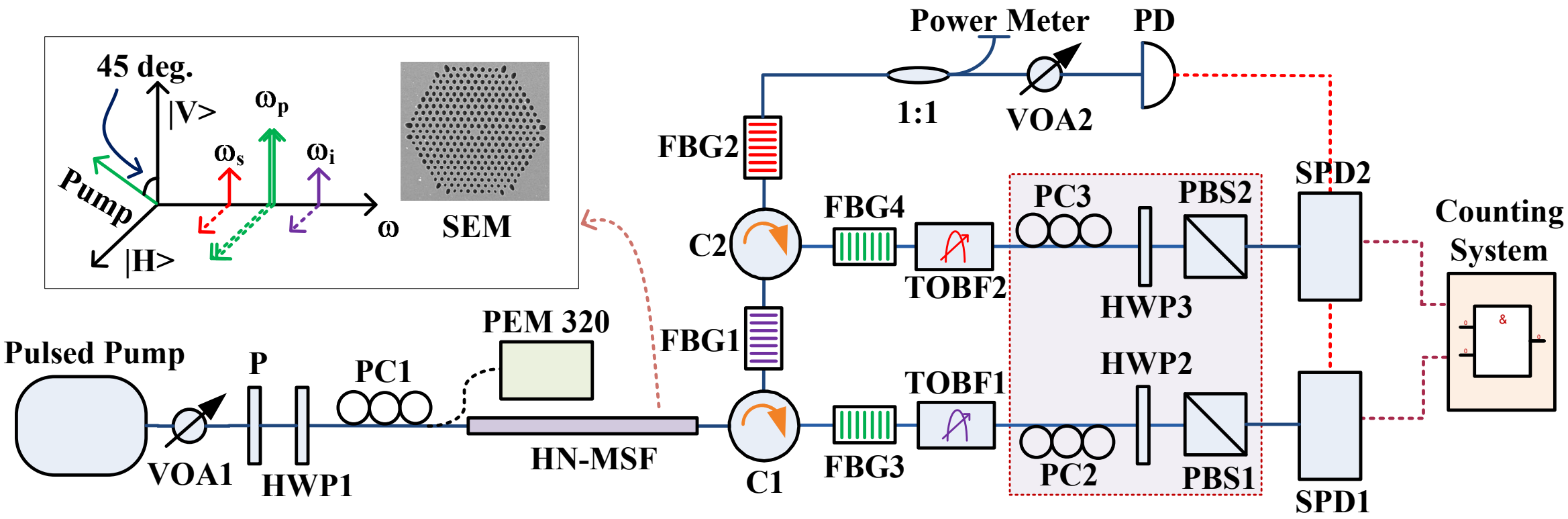

Fig.1 (Color online) Experiment setup

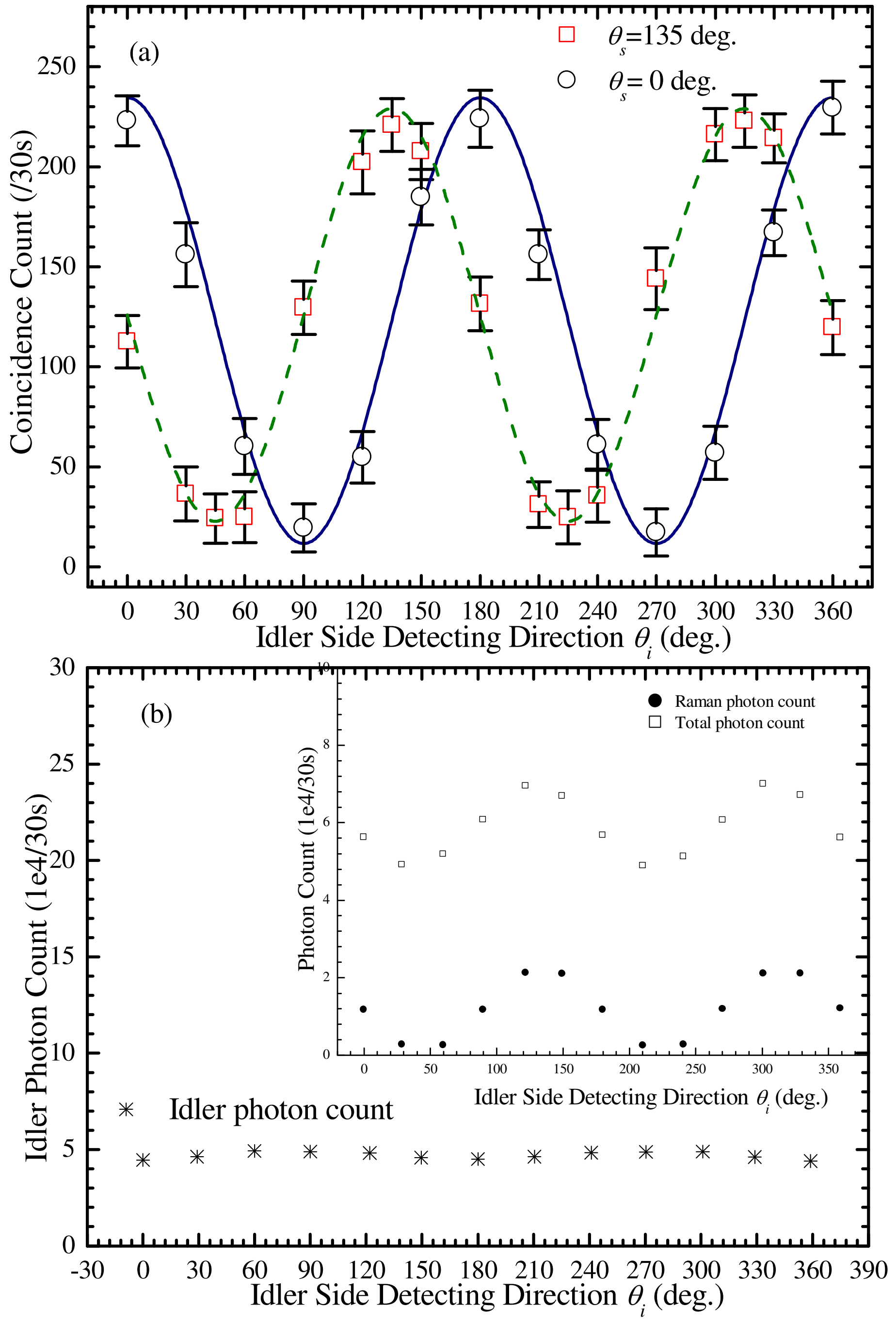

Fig.2 (Color online) Experiment results of polarization entanglement demonstration, (a) coincidence count under different $\theta_i$, (b) idler side photon count under different $\theta_i$.

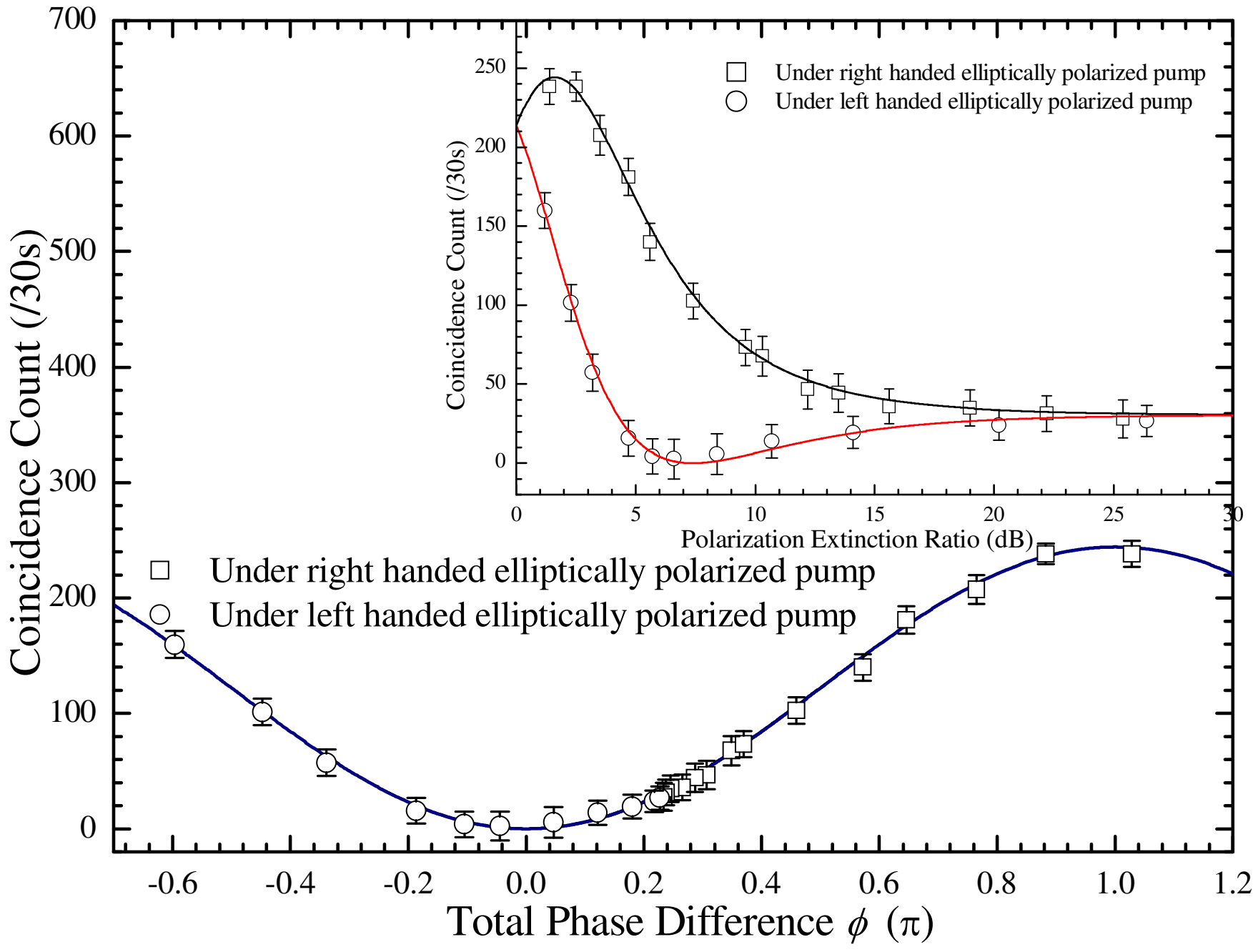

Fig.3 (Color online) Polarization-entangled Bell states generation by controlling the pump polarization state